\documentclass[%
 amsmath,amssymb,
 aip,
]{revtex4-1}
\usepackage{graphicx}% Include figure files
\usepackage{dcolumn}% Align table columns on decimal point
\usepackage{bm}% bold math
\usepackage{colortbl}
\usepackage{hyperref}% add hypertext capabilities
\newcolumntype{R}[1]{>{\raggedleft\arraybackslash }b{#1}}
\newcolumntype{L}[1]{>{\raggedright\arraybackslash }b{#1}}
\newcolumntype{C}[1]{>{\centering\arraybackslash }b{#1}}

\begin{document}

%\preprint{APS/123-QED}

\title{Control of magnon-photon coupling strength in a planar resonator/YIG thin film configuration}

%--------------------------------Author------------------------------------------------------------------------------------------------------------------
%-------------------------------------------------------------Author-------------------------------------------------------------------------------------
\author{V. Castel}
% \email{vincent.castel@telecom-bretagne.eu}
\affiliation{ 
T\'el\'ecom Bretagne, Technopole Iroise-Brest, CS83818, 29200 Brest, France.%\\This line break forced with \textbackslash\textbackslash
}%
\author{A. Manchec}
\affiliation{Elliptika (GTID), 29200 Brest, France.%\\This line break forced with \textbackslash\textbackslash
}%
\author{J. Ben Youssef}
\affiliation{ Universit\'e de Bretagne occidentale, Laboratoire de Magn\'etisme de Bretagne CNRS, 29200 Brest, France%\\This line break forced with \textbackslash\textbackslash
}%

\date{\today}% It is always \today, today,

\begin{abstract}
A systematic study of the coupling at room temperature between ferromagnetic resonance (FMR) and a planar resonator is presented. The chosen magnetic material is a ferrimagnetic insulator (Yttrium Iron Garnet: YIG) which is positioned on top of a stop band (notch) filter based on a stub line capacitively coupled to a 50 $\Omega$ microstrip line resonating at 4.731 GHz. Control of the magnon-photon coupling strength is discussed in terms of the microwave excitation configuration and the YIG thickness from 0.2 to 41 $\mu$m. From the latter dependence, we extract a single spin-photon coupling of g$_{0}$/2$\pi$=162$\pm$6 mHz and a maximum of an effective coupling of 290 MHz.

%\begin{description}
%72.25.Ba, 72.25.Pn, 75.78.-n, 76.50.+g
%\end{description}
\end{abstract}


\maketitle

\section{Introduction}

A recent field, known as cavity spintronics\cite{Flatte2010, DawnCavity}, is emerging from the progress of spintronics combined with the advancement in Cavity Quantum Electrodynamics (QED) and Cavity Polaritons\cite{Sanders1978,Mills}. Cavity QED allows the use of coherent quantum effects for quantum information processing and offers original possibilities for studying the strong interaction between light and matter in a variety of solid-state systems\cite{Cohen-Tannoudji2004, Laflamme2002, Wallraff2004}. A superconducting two-level system is quantum coherently coupled to a single microwave photon and an analogy to spintronics (spin two-level system) has been made. The high spin density of the ferromagnet used in Ref.\cite{Huebl2013, Tabuchi2014} has made it possible to create a strongly coupled magnon mode. Magnon-photon coupling has been investigated in several experiments at room temperature where a microwave resonator (three-dimensional cavity\cite{Kang2008,Gollub2009,Zhang2014,Lambert2015,Haigh2015,Bai2015,Maier-Flaig2016,Bai2016,Harder2016} and planar configuration\cite{Stenning2013,Bhoi2014,Klingler2016}) was loaded with a ferrimagnetic insulator such as the Yttrium Iron Garnet (YIG, thin film and bulk). A study on a transition metal like Py (structured thin film) coupled with a Split Ring Resonator (SRR) was done by Gregory et al.\cite{Gregory2014} in order to demonstrate the possibility to achieve YIG-type functionalities and to overtake the working frequency limitation of YIG. More recently, L. Bai et al.\cite{Bai2015} have developed an electrical method to detect magnons coupled with photons. This method has been established by placing a hybrid YIG/Pt system in a microwave cavity showing distinct features not seen in any previous spin pumping experiments but already predicted by Cao et al.\cite{Cao2014}.

\section{Compact design description}
The main objective of the present paper is to demonstrate the control of a magnon-photon coupling regime at room temperature in a compact design based on a stub line coupled with a microstrip with YIG thin film. Control of a magnon-photon coupling regime in such configuration offers manifold opportunities in the development of integrated spin-based microwave applications, such as a sensitive reconfigurable stop-band filtering function.

\begin{figure}
	\includegraphics[width=9.5cm]{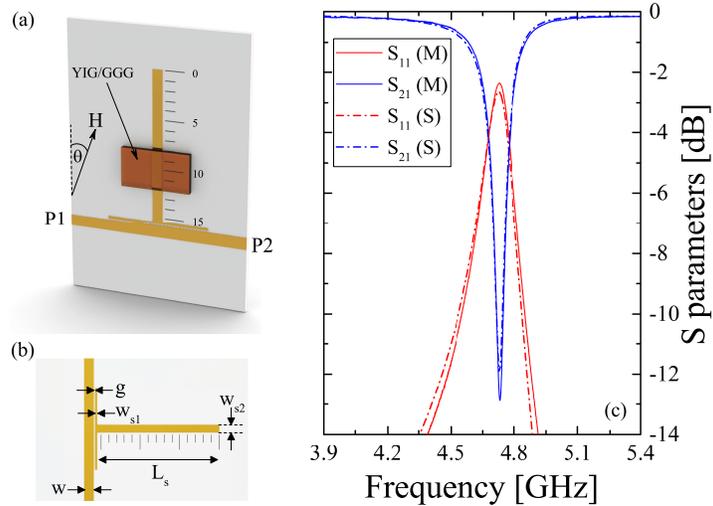}% Here is how to import EPS art
	\caption{\label{fig:Fig1} 
		(a) Experimental setup: A Vector Network Analyzer (VNA) is connected to a 50 $\Omega$ microstrip line which is capacitively coupled to the resonator. Numbers from 0 to 15 are referred to the YIG sample position (an example is given for which the center of the sample is placed at x=10 mm=0.25 $\lambda$). (b) Dimension of the microwave stop band resonator configuration. (c) Frequency dependence of S parameters measured (M) and simulated (S) by CST simulation of the empty resonator (without YIG sample).
	}
\end{figure}

Instead of using a Split Ring Resonator (SRR) configuration, the choice was made to study the YIG thickness dependence of the coupling regime with a stub line geometry for which the microwave excitation of the magnetic medium is simplified. Figure \ref{fig:Fig1} (a) represents the sketch of the experimental setup based on the stub/YIG film system excited by a microwave signal under an in-plane static magnetic field, H, at $\theta$=0$^{\circ}$. $\theta$ is defined by the angle formed between H and the stub line. P1 and P2 correspond to the 2 ports of the VNA for which a TSOM calibration were realized (including cables). The frequency range is fixed from 3 to 6 GHz at an microwave power of P=-10 dBm. The circuit is fabricated on a pre-metallized (double-sided 25 $\mu$m copper coating) ROGERS substrate (3003) presenting a relative permittivity of $\varepsilon_{r}$=3 and losses tan$\delta$=2$\times$10$^{-3}$ (dimension are shown in Fig. \ref{fig:Fig1} (b)). The narrow stop-band (notch) resonator configuration is based on a main 50 $\Omega$ microstrip line (W=1.23 mm) coupled by a gap of g=150 $\mu$m to an open circuited half wavelength stub (L$_{\textrm{s}}$=15.52 mm, W$_{\textrm{s1}}$=280 $\mu$m and W$_{\textrm{s2}}$=1.0 mm). It has been designed with an attenuation of 13 dB at 4.75 GHz and 80 MHz of bandwidth. The frequency dependence of the S11 and S21 parameters (measured and simulated) of the empty resonator are shown in Fig. \ref{fig:Fig1} (c). The S21 resonance peak has a half width at half maximum (HWHM) $\Delta$F$_{\textrm{HWHM}}$ of 32 MHz indicating that the damping of the resonator (working at the frequency $F_{0}$) is $\beta$=$\Delta$F$_{\textrm{HWHM}}$/$F_{0}$=1/2Q=6.8$\times$10$^{-3}$. This leads to a quality factor Q of 74. The latter definitions of the Q factor and $\beta$ (used in the following discussion) are based on the expression extracted from recent studies in the field of magnon-photon coupling (2D\cite{Bhoi2014,Klingler2016} and 3D cavities\cite{Bai2015,Maier-Flaig2016,Bai2016,Harder2016,Lambert2015,Haigh2015}). Nevertheless, this definition does not reflect properly the electrical performances of our notch filter which are defined by Q$_{0}=F_{0}/\left[\Delta F^{S21}_{-3dB}(1-S11_{F_{0}}) \right]$=153.
%\IEEEpubidadjcol

Single-crystal Y$_3$Fe$_5$O$_{12}$ (YIG) samples from 0.2 to 41 $\mu$m were elaborated by Liquid Phase Epitaxy (LPE) on top of a 500 $\mu$m thick GGG substrate in the (111) orientation. YIG samples have been cut in rectangular shape (4 mm$\times$7 mm) and placed on the stub line as shown in Fig. \ref{fig:Fig1} (a) with the crystallographic axis [1,1,$\bar{2}$] parallel to the planar microwave field generated by the stub line. The magnetic losses (Gilbert damping parameter, $\alpha$) of the set of YIG sample were investigated by FMR measurements using a highly sensitive wideband resonance spectrometer within a range of 4 to 20 GHz. Measurements were carried out at room temperature with a static magnetic field applied in the plane of YIG samples. These characterizations have given rise to a parameter $\alpha\leq$2$\times$10$^{-4}$ for the set of samples which is in agreement with previous studies \cite{Castel2014}.

\section{Results and discussion}

\begin{figure}[ht]
	\includegraphics[width=10cm]{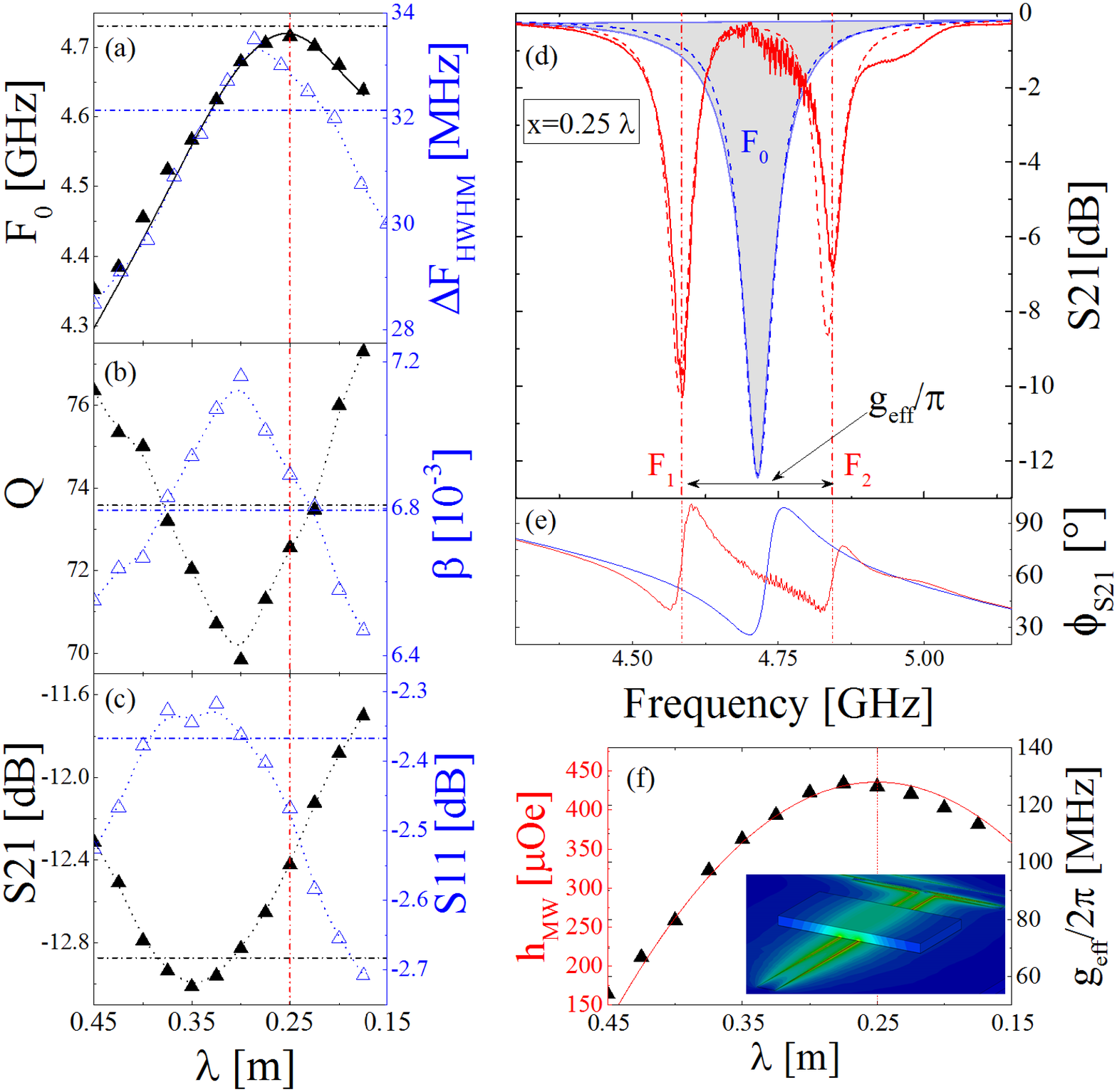}% Here is how to import EPS art
	\caption{\label{fig:Fig2} 
		(a) to (c): YIG sample position dependence (at H=0 Oe) of (a) $F_{0}$ and $\Delta$F$_{\textrm{HWHM}}$, (b) Q factor and losses, (c) S11 and S21 at the resonant frequency $F_{0}$. Horizontal dash dot lines correspond to parameters extracted from the empty resonator whereas the vertical dash dot line illustrates the position of YIG (respect to the center) at x=0.25 $\lambda$ as shown in Fig. \ref{fig:Fig1} (a). (d) and (e): Signature of the coupling. Frequency dependence of the magnitude in dB of S21 (d) and the associated phase $\phi_{\textrm{S21}}$ (e) for a YIG position at x=0.25 $\lambda$. Solid blue and red line correspond respectively to the response at H=0 Oe and at H=H$_{\textrm{RES}}$. $F_{1}$ and $F_{2}$ (represented by vertical red dash lines) correspond to hybridized mode frequencies whereas g$_{\textrm{eff}}$/2$\pi$ corresponds to the coupling strength parameter. (f) Dependence of g$_{\textrm{eff}}$/2$\pi$ (measured, black triangles) and microwave field amplitude (CST simulation, solid red line) as function of the YIG sample position. All measurements have been carried on at room temperature on a YIG sample which presents a thickness of (9 $\mu$m). The inset shows the spatial distribution of the microwave magnetic field simulated at $F_{0}$.
	}
	
\end{figure}

We first studied the magnitude of the coupling strength as a function of the position of a 9 $\mu$m YIG sample on top of the planar resonator, as shown in Fig. \ref{fig:Fig1} (a). Figures \ref{fig:Fig2} (a) to (c) illustrate the dependence of the resonator features (at H=0 Oe), such as the resonant frequency $F_{0}$, linewidth $\Delta$F$_{\textrm{HWHM}}$, damping of the resonator $\beta$ (Q factor), and the dependence of S11 and S21 at the resonant frequency $F_{0}$. $F_{0}$ can be tuned from 4.35 to 4.715 GHz (tuning of 8.4 \%) and presents a maximum at x=0.25 $\lambda$ which is closed to the resonant frequency of the empty resonator (represented by horizontal dash dot lines). Note that the YIG position corresponds to the center of the sample as illustrated in Fig. \ref{fig:Fig1} (a). The wavelength is defined by $\lambda=\frac{\lambda_{0}}{\sqrt{\varepsilon_{eff}}}$, where $\varepsilon_{eff}$ corresponds to the effective permittivity. The configuration of the notch filter (open circuit (OC) at x=0.5 $\lambda$) induced necessarily the definition of short circuit (SC) at x=0.25 $\lambda$ which explains the limited impact of YIG (at this position) on the resonator features. Introduction of a YIG layer ($\varepsilon_{r}$=15 and losses tan$\delta$=2$\times$10$^{-4}$) on CST simulation make it possible to correctly reproduce this dependence at zero field (solid black line) which is attributed to the modification of the effective permittivity. Here, only the electrical properties of the YIG sample were taken into account. Contrary to a ferromagnetic conductor, such as an extended thin film of Permalloy (Py, NiFe)\cite{Gregory2014}, no eddy current shielding effect of YIG on the stub line was observed. YIG is a ferrimagnetic insulator with a band gap of 2.85 eV and the high quality YIG samples used in this study allow the reduction of negative impacts on the planar resonator. As shown in Fig. \ref{fig:Fig2} (a), $\Delta$F$_{\textrm{HWHM}}$ reaches a maximum of 34 MHz at x=0.3$\lambda$, which represents an enhancement of only 2 MHz with respect to the empty resonator (a reduction of 5 MHz being achievable at x=0.45$\lambda$). Slightly changes in the Q factor ($\beta$) from 70 to 77 (6.4 to 7.2$\times$10$^{-3}$) were obtained. In the meantime, attenuation represented by the S11 parameter at $F_{0}$ are closed to the value extracted from the empty resonator from x=0.4 to 0.3$\lambda$ (-2.35 dB). 

For each position of the YIG sample, measurement at room temperature of the frequency dependence of S parameters (magnitude and phase) at P=-10 dBm was done with respect to the applied magnetic field. Figure \ref{fig:Fig2} (d) and (e) represent, respectively, the frequency dependence of S21 and $\Phi_{\textrm{S21}}$ of the notch/YIG system at x=0.25 $\lambda$ (as shown in Fig. \ref{fig:Fig1} (a)). Solid (dash) blue and red lines are associated with the experimental (analytic solution from Eq. (3) from Ref.\cite{Harder2016}) response under an applied magnetic field of H=0 Oe and H=H$_{\textrm{RES}}$, respectively. The FMR and the notch filter interact by mutual microwave fields, generated by the oscillating currents in the stub and the FMR magnetization precession which led to the following features observed in Fig. \ref{fig:Fig2} (d) and (e): (i) Hybridization of resonances (magnitude and phase\cite{Harder2016phase}), (ii) Annihilation of the resonance at $F_{0}$, and (iii) Generation of two resonances at $F_{1}$ and $F_{2}$. At the resonant condition H=H$_{\textrm{RES}}$, the frequency gap, $F_{\textrm{gap}}$, between $F_{1}$ and $F_{2}$ is directly linked to the coupling strength of the system ($F_{\textrm{gap}}$/2=g$_{\textrm{eff}}$/2$\pi$). Several models can be used to analyze the hybridized mode frequency $F_{1}$ and $F_{2}$ in the system. Recently, Harder et al. \cite{Harder2016} have examined the accuracy to describe the microwave transmission line shape of a cavity/YIG system through three different models: coupled harmonic oscillators, dynamic phase correlation, and microscopy theory. Here, the analysis has been focussed on the harmonic coupling model for which we can define the upper ($F_{2}$) and lower ($F_{1}$) branches by:

\begin{equation}
F_{1,2}=\dfrac{1}{2}\left[ \left( F_{0}+F_{r}\right) \pm \sqrt{\left( F_{0}-F_{r}\right)^{2}+k^{4}F_{0}^{2}}\right] 
\label{F1F2} 
\end{equation}  
The FMR frequency, $F_{r}$, is modelled by the Kittel equation, $F_{r}=\dfrac{\gamma}{2\pi}\mu_{0}\sqrt{H(H+M_{s})}$, which describes the precession frequency of the uniform mode (without taking into account spin wave distribution) in an in-plane magnetized ferromagnetic film. The parameter $k$ used in Eq. \ref{F1F2} corresponds to the coupling strength which is linked to the experimental data g$_{\textrm{eff}}$/2$\pi$ by the following equation\cite{Harder2016}: $F_{\textrm{gap}}=F_{2}-F_{1}=k^{2}F_{0}$. As shown in Fig. \ref{fig:Fig2} (f), sensitive control of g$_{\textrm{eff}}$/2$\pi$ (and thus $k$) can be achieved by adjusting the YIG sample position on the resonator from 54 MHz ($k$=0.1573) at x=0.45 $\lambda$ to 127 MHz ($k$=0.2315) at x=0.25 $\lambda$. In order to understand the dependence of g$_{\textrm{eff}}$/2$\pi$ on the YIG position, CST simulations were carried out in order to determine the microwave field ($h_{\textrm{{\tiny MW}}}$) generated at each position (represented in solid red line). It ends up that $h_{\textrm{{\tiny MW}}}$ follows exactly the same trend of the coupling factor, in agreement with the fact that the effective coupling strength depends on the mutual microwave field interaction between the FMR and the stub line. The latter dependence is defined by the following equation \cite{Zhang2014, Tabuchi2014}:
\begin{equation}
\frac{g_{eff}}{2\pi}=\frac{\eta}{4\pi}\gamma_{e}\sqrt{\frac{\hbar\omega_{0}\mu_{0}}{V_{c}}}\sqrt{N},
\label{Gdep} 
\end{equation}
where $\gamma_{e}$ is the electron gyromagnetic ratio of 2$\pi\times$28.04 GHz/T, $\mu_{0}$ is the 
permeability of the vacuum, $V_{c}$ corresponds to the volume of the cavity, and $N$ is the total number of spins. The coefficient $\eta\leq 1$ describes the spatial overlap and polarization matching conditions between the microwave field and the magnon mode. In agreement with Zhang et al. \cite{Zhang2014} (Appendix A), we demonstrated the dependence of g$_{\textrm{eff}}$/2$\pi$ as function of the spatial distribution of the microwave magnetic field along the stub line which is maximum at x=0.25 $\lambda$ (short circuit).

\begin{table}[h]
	\caption{Notch/YIG configuration versus SRR \& cavity/YIG systems}
	\label{table1}
	\centering
	\begin{tabular}{l||c|c|c|c}	
		Ref.& $\beta$ [10$ ^{-3} $] & g$_{\textrm{eff}}$/2$\pi$ [MHz]& $F_{0}$ [GHz] & $k$  \\
		
		\hline
		& & & &  \\ [-6.5pt]
		\cite{Bai2015} &1.8 &80 & 10.506  &0.1234\\ 
		\cite{Maier-Flaig2016} &0.708 & 31.8 & 9.65 &0.0812\\
		\cite{Bai2016} &2.3 & 65 & 10.847 &0.1095  \\
		\cite{Harder2016} &0.3 &31.5 &10.556 &0.0773 \\
		\cite{Lambert2015,Haigh2015} &1.92 &130 &3.535 &0.2712\\
		\cite{Bhoi2014} &9.85 &270 &3.2 &\textbf{0.4108}\\
		\cite{Klingler2016} &5.04 &63 &4.96 &0.1594\\
		This work$^{[1]}$ &6.89  &127 & 4.716 &0.2315\\	
		This work$^{[2]}$ &6.89 &\textbf{290} & 4.719 &0.3508\\
	\end{tabular}
\end{table}

Table \ref{table1} gives a picture of recent work on the determination and control of magnon-photon coupling regimes in SRR\cite{Bhoi2014,Klingler2016} and cavity\cite{Bai2015,Maier-Flaig2016,Bai2016,Harder2016,Lambert2015,Haigh2015}/YIG systems. Ref. \cite{Bai2015,Maier-Flaig2016,Bai2016} correspond to the research field associated with the electrical detection of magnons coupled with photons via combined phenomena in a hybrid YIG/Pt system. Despite the fact that these later studies have been realized in a cavity, insertion of a hybrid stack including a highly electrical conductor induced an enhancement of the intrinsic loss rate $\beta$ (factor of 5\cite{Bai2015} to 12\cite{Maier-Flaig2016}). The value of $k$ obtained at the optimized position at x=0.25 $\lambda$ is significantly higher than Ref.\cite{Bai2015,Maier-Flaig2016,Bai2016,Harder2016} and comparable to Ref.\cite{Lambert2015,Haigh2015,Klingler2016} but still much smaller than the value obtained by Bhoi et al.\cite{Bhoi2014}. It should be noted that the normalization of $k$ by the intrinsic loss rate $\beta$ changes the latter comparison drastically.

\begin{figure}[h]
	\includegraphics[width=9cm]{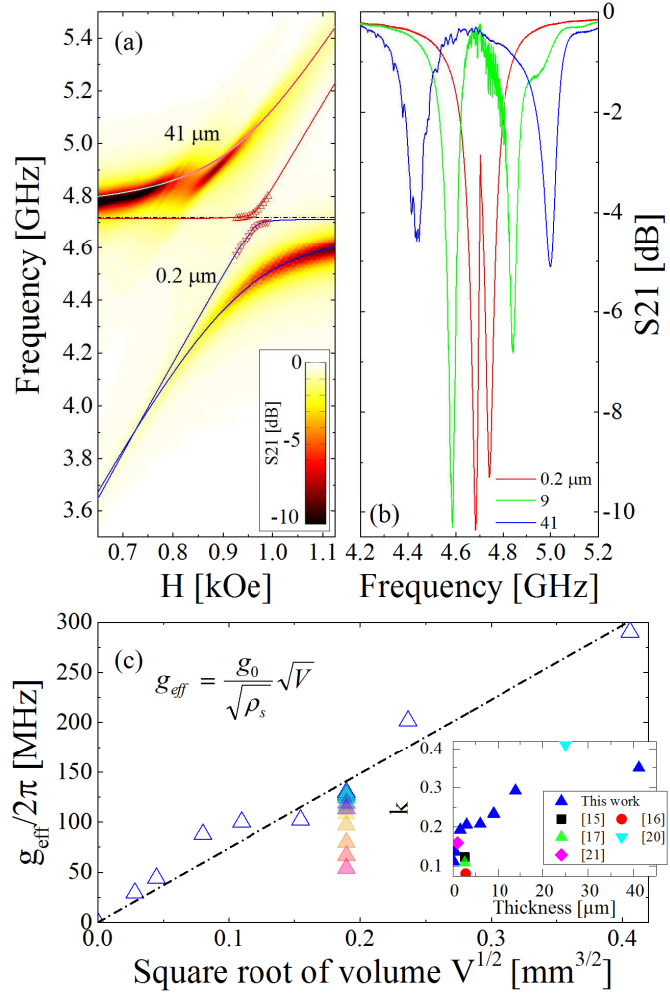}% Here is how to import EPS art
	\caption{\label{fig:Fig3} 
		Control of the coupling strength as function of the YIG thickness. (a) Magnetic field dependence of the frequency: Observation of the strong coupling regime via the anti-crossing fingerprint. The color map is associated to the response of the thicker YIG sample (41 $\mu$m). (b) Frequency dependence of S21 at the resonant condition for various YIG thickness (0.2, 9, and 41 $\mu$m). (c) Coupling strength of the Kittel mode to the microwave resonator mode as a function of the square root of the YIG volume. Colored triangles correspond to the dispersion of g$_{\textrm{eff}}$/2$\pi$ from Fig. \ref{fig:Fig2} (f). The inset represents the YIG thickness dependence of $k$. All measurements were done at room temperature and at x=8 mm as shown in Fig. \ref{fig:Fig1} (a).
	}
\end{figure}

Next, the dependence of the coupling strength between the FMR and the notch filter was investigated with respect to the YIG thickness from 0.2 to 41 $\mu$m. YIG samples were placed at the optimized position which has been determined previously (x=0.25 $\lambda$). This particular position gives an access to the highest coupling (determined at P=-10 dBm) and presents the best compromise in terms of the electrical performance of the notch filter. Sample position was adjusted by tracking $F_{0}$ at H=0 Oe ($F_{0}$=4.715$\pm$ 0.002 GHz). It should be noted that no dependence of the insertion rate of the resonator ($\beta$=6.89$ \pm$0.01 $10^{-3}$) and attenuation (S11$_{F_{0}}$=-2.45$ \pm$ 0.04 dB) have been observed with respect to the YIG thickness. As shown in Fig. \ref{fig:Fig3} (a), we demonstrated a strong coupling regime via the anti-crossing fingerprint. A good agreement of $F_{1,2}$ based on Eq. \ref{F1F2} (solid lines) with experimental data is obtained for the various YIG thicknesses. The color plot in Fig. \ref{fig:Fig3} (a) is associated with the S21 parameter for which the dark area corresponds to a magnitude of -10 dB. This representation underlines the complexity of the response by increasing the YIG thickness from 0.2 to 41 µm, well illustrated by the additional anti-crossing signature between 0.75 and 0.90 kOe (upper resonance). In the following discussion, the extraction of the coupling factor is only based on the uniform mode without taking into account the dispersion relation of spin waves. Figure \ref{fig:Fig3} (b) represents the frequency dependence of the transmission spectra for the notch/YIG system at the resonant condition for which the effective coupling was extracted. Control of the frequency gap can be achieved from 59 to 581 MHz through an enhancement of the YIG thickness from 0.2 to 41 $\mu$m, respectively. Parameters associated with the thicker YIG are summarized in Tab \ref{table1} (last row). 

The originality of this study is described in Fig. \ref{fig:Fig3} (c) which represents the dependence of the effective coupling g$_{\textrm{eff}}$/2$\pi$ as a function of the square root of the YIG volume interacted with the 1 mm width microwave resonator (V=4 mm$\times$1 mm$\times$YIG$_{\textrm{{\tiny thick}}}$ $\mu$m). Cao et al.\cite{Cao2014} shows that the filling factor of magnetic medium in a cavity can be used as a measure of the total number of spins, $N$. The large effective coupling strength is due to the large spin density of YIG, $\rho_{s}$=2.1$\times$10$^{22}$ $\mu_{\textrm{B}}$cm$^{-3}$ ($\mu_{\textrm{B}}$; Bohr magneton). The linear fit of the dependence presented in Fig. \ref{fig:Fig3} (c) gives rise to a slope of 742 $\pm$29 MHz mm$^{3/2}$ which makes it possible to extract the single spin-photon coupling g$_{0}$/2$\pi$=162$\pm$6 mHz based on the following equation\cite{Maier-Flaig2016,Tabuchi2014} g$_{\textrm{eff}}$=g$_{0}\sqrt{N}$. Tabuchi et al.\cite{Tabuchi2014} demonstrate a good agreement between the evaluation of the single-spin coupling strength from the fitting (39 mHz) and theory derived from the quantum optics community (38 mHz). The latter value is calculated from Eq.\ref{Gdep} by taking the coefficient $\eta=1$. The higher value of g$_{0}$/2$\pi$ obtain in the present work is mainly due to the compactness of our resonator. A rough estimation of the volume of our notch filter working at $F_{0}$=4.750 GHz can be done by using a one dimensional transmission-line cavity\cite{Schoelkopf2008} which defines the volume as $V_{c}=\pi r^{2}\lambda/2$. By assuming $r$=0.5 mm (distance between the feed line and the ground), $\lambda=c/(\sqrt{\varepsilon_{eff}}F_{0})$, and $\eta=1$, we evaluate g$_{0}$/2$\pi$=177 mHz when the cavity is completely filled by air ($\varepsilon_{eff}=\varepsilon_{r}$=1) which is closed to the extracted value of g$_{0}$/2$\pi$ from the fitting. Nevertheless, this value does not reflect the fact that the cavity is nonuniformly filled with other dielectric materials such as the substrate, GGG, and YIG. An enhancement of g$_{0}$/2$\pi$ from 177 to 219 mHz can be achieved by taking into account an effective permittivity of $\varepsilon_{eff}$=2.449. The latter value is determined\cite{EpsEFF} for the notch filter loaded with a YIG film at x=0.25 $\lambda$ which induced a diminution of $F_{0}$ from 4.750 to 4.715 GHz.

%The comparison with Ref.\cite{Tabuchi2014} is not obvious. First, our 2D resonator generates a inhomogeneous microwave field distribution which induced the excitation of backward volume magnetostatic spin wave (BVMSW) through the out-of-plane component of $h_{\textrm{{\tiny MW}}}$. Second, the magnetization precession cone angle at resonance is a missing parameter in the analysis. Nevertheless, the much higher microwave power used in our experiment (-10dBm compared to only -123 dBm) partially explained the higher value of g$_{0}$/2$\pi$ for our system (4 times higher than Ref.\cite{Tabuchi2014}).

 %------------------------------------------------------------------------------------------------------------------------------------------------------------------
%------------------------------------------------------------------------------------------------------------------------------------------------------------------
%------------------------------------------------------------------------------------------------------------------------------------------------------------------

We have demonstrated the presence of a strong coupling regime via the anti-crossing fingerprint of the FMR from an magnetic insulator and a planar resonator. Control of the coupling with respect to the YIG thickness from 0.2 to 41 $\mu$m makes it possible the determination of the single spin photon coupling of our system at room temperature. We have found that g$_{0}$/2$\pi$ in a thin film configuration is equal to 162$\pm$6 mHz. In the meantime, we demonstrate an effective coupling strength of 290 MHz for the thicker YIG. Improvement on insertion losses of the planar resonator can be achieved in order to be more competitive regarding the 3D cavity system by changing the design of the resonator (SRR, array of SRR, enhancement of the capacitive coupling) and/or by using a low loss substrate ($<10^{-2}$). Other tuning channel of the coupling strength such as microwave power or spin wave dispersion might help for the realization of YIG-based devices.

%------------------------------------------------------------------------------------------------------------------------------------------------------------------
%------------------------------------------------------------------------------------------------------------------------------------------------------------------
%------------------------------------------------------------------------------------------------------------------------------------------------------------------

%\nocite{*}
\bibliography{SpinCavity}% Produces the bibliography via BibTeX.

\end{document}